\documentstyle[12pt]{article}
\newcommand{\be}{\begin{equation}}
\newcommand{\ee}{\end{equation}}
\newcommand{\bea}{\begin{eqnarray}}
\newcommand{\eea}{\end{eqnarray}}
\newcommand{\sptwo}{1.4}
\newcommand{\doublespace}{\edef\baselinestretch{\sptwo}\Large\normalsize}

\begin{document}
\hspace*{\fill} PURD-TH-94-01\\
\hspace*{\fill} VAND-TH-94-01\\
\vspace{0.5in}

\begin{center}
{\large\bf MASS BOUNDS IN THE STANDARD MODEL\\}
\end{center}
~\\
\begin{center}
{\bf T.E. Clark, B. Haeri, S.T. Love, M.A. Walker}\\
{\it Department of Physics}\\ 
{\it Purdue University}\\
{\it West Lafayette, IN 47907-1396}
~\\
~\\
{and}
~\\
~\\
{\bf W.T.A. ter Veldhuis}\\
{\it Department of Physics and Astronomy\\
Vanderbilt University\\
Nashville, TN 37235}
\end{center}
~\\
~\\
\begin{center}
{\bf Abstract}
\end{center}

Nonperturbative triviality and vacuum stability mass 
bounds are obtained for 
the Higgs scalar and top quark degrees of freedom in the standard electroweak 
model using Wilson 
renormalization group techniques. Particular attention is given to the effect 
of the generalized top Yukawa coupling on the scalar mass upper bound.
\pagebreak

\doublespace

The scalar sector of the standard electroweak model 
which is responsible for the electroweak symmetry breaking is 
characterized by two parameters.  The dimensionful scalar mass term coupling 
is phenomenologically set by the electroweak symmetry breaking scale, while the 
dimensionless quartic scalar self coupling is related to the magnitude of the 
physical Higgs scalar mass. While this mass is completely unconstrained at 
tree level, the radiative corrections lead to inconsistencies 
unless the mass is constrained to be less than some upper bound.  This 
so-called triviality bound is a reflection of the infrared 
Gaussian (trivial) fixed point of the model which, in turn, 
can be traced to the non-asymptotically free nature 
of the scalar self coupling.  Its increasing growth tends to destabilize the 
ultraviolet physics unless suitably constrained.  

The triviality bound has been calculated in 
perturbation theory$^{[1]}$ and found to produce a scalar mass upper 
bound of O(1 TeV).  However, since such a scalar mass corresponds to 
a nonperturbative value of the scalar self coupling, the result requires 
further scrutiny via calculation within a nonperturbative framework.  Various 
nonperturbative studies ranging from lattice simulations$^{[2]}$ to 1/N 
expansions$^{[3]}$ 
have been carried out for the self coupled scalar system.  Another approach 
advocated by Hasenfratz and Nager$^{[4]}$ employs a nonperturbative Wilson 
renormalization group equation$^{[5]}$. In each case, the perturbative result 
has been substantiated and an upper bound Higgs scalar mass of O(1 TeV) has 
again been obtained. The basic equality of these bounds can be traced to the 
presence of the infrared Gaussian fixed point in all analyses. 

Of course, when considering the standard model, the scalar sector should not 
be treated in isolation.  Since the Higgs mass bound arises from constraints on 
the ultraviolet physics where the gauge couplings are all small, their effect 
can be adequately accounted for within perturbation theory.  The scalar 
doublet also couples to the various fermions via Yukawa 
couplings whose magnitude is set by the associated fermion mass.  Since these 
couplings are also non-asymptotically free, the self consistency of the model 
likewise forces them to be appropriately constrained.  All the presently 
observed fermions are sufficiently light that the effects of their Yukawa 
coupling are also amenable to a perturbative study.  On the other hand, the 
present experimental bound on the top quark mass is such that its Yukawa 
coupling may require a nonperturbative analysis.

In the present note, we determine the allowed range of Higgs scalar and top 
quark masses using a Wilson renormalization group equation (WRGE) approach.  
Such an equation containing both scalars and fermions was obtained 
previously$^{[6]}$ by three of the authors and can be applied here.  The WRGE 
nonperturbatively relates the form of the Euclidean action at a scale 
$\Lambda (t) = e^{-t} \Lambda$ to the action at scale $\Lambda$ for $t>0$. 
It is derived by demanding that the 
physics, ie. correlation functions, remain unchanged as degrees of freedom 
carrying momentum between the scales $\Lambda$ and $e^{-t} \Lambda$ are 
integrated out. Thus either action can be used to equivalently describe the 
physics on all scales less than $e^{-t} \Lambda$ and both actions lie on the 
same Wilson renormalization group trajectory. The lower scale action is 
constructed by appropriately changing the coefficients of the operators 
already present 
in the scale $\Lambda$ action and by introducing new operators which were not 
present at scale $\Lambda$ in such a way so as to keep the physics unchanged. 

In our previous work$^{[6]}$, we 
applied the equation to study a Higgs-Yukawa model of a Dirac fermion 
interacting with a 
real scalar in a way which was invariant under a discrete $\gamma_5$ symmetry 
which prevented an explicit fermion mass term. Since one of the major 
attributes of 
the formalism is that it can also be used to describe chiral fermions, 
we can also employ the same equation to study the standard model 
which contains chiral fermions interacting with a complex electroweak doublet 
of scalar fields in a $SU(2)_L \times U(1)$ invariant manner.  Since the gauge 
interactions can be handled perturbatively, we do not have to use the full 
WRGE structure in their treatment. This is fortuitous since the WRGE 
formalism is not easily adaptable to deal with strong gauge couplings.  
Unfortunately, just as in the discrete symmetry case, to render the WRGE 
tractable to analysis even in the absence of gauge interactions requires the 
introduction of various approximations.  
The first of these is a local action approximation$^{[4,6]}$ which 
neglects 
anomalous dimensions and derivative interactions and thus constitutes a first 
term in a momentum expansion.  The WRGE then reduces to an equation for a 
generalized potential function $U(\rho, Y; t)$,  where $\rho = 2 \varphi^
{\dagger} \varphi$ and $Y = \sqrt{2} ( \bar{t}_L t_R \varphi_0 + 
\bar{b}_l t_R 
\varphi_- + \bar{t}_R b_L \varphi_+ + \bar{t}_R t_L \varphi_0^{\dagger})$.  
Here we have assumed that all the fermion dependence of $U$ appears as a 
function of $Y$ which is the $SU(2) \times U(1)$ invariant coupling of the 
scalar doublet $\phi^{\dagger} = (\varphi_0^{\dagger}, \varphi_+)$ and its 
conjugate to the top 
and bottom quarks, $t, b$, respectively.  So doing, we have neglected possible 
dependences on the fermions appearing in higher dimensional (eg. four-fermion) 
$SU(2) \times U(1)$ invariant structures.
A further truncation, made solely 
to facilitate the subsequent numerical analysis, is to retain terms in $U$ 
only up to linear in $Y$, but still multiplying a general function of $\rho$.  
We thus write 
\be
U(\rho, Y; t) = V(\rho, t) + Y G(\rho, t)
\ee
as the sum of an effective potential $V(\rho, t)$ which determines the vacuum 
structure and a generalized Yukawa interaction $YG(\rho, t)$.  Defining 
$F(\rho, t) = V_{\rho}(\rho, t) = \frac{\partial V(\rho, t)}{\partial \rho}$, 
(subscripts denoting 
differentiation) so that the ground state corresponds to the zeroes of $F$, 
the WRGE$^{[7]}$ reduces to the two coupled partial differential equations 
\bea
{\partial F\over \partial t} &=& \frac{1}{16\pi^2}\left[ \frac{6F_\rho}{1+2F}
+\frac{6F_\rho +4\rho F_{\rho\rho}}{1+2F+4\rho F_\rho}-\frac{12G^2+ 24\rho 
GG_\rho}{1+\rho G^2}\right] +2F-2\rho F_\rho \nonumber\\
{\partial G\over \partial t} &=& \frac{1}{16\pi^2}\left[\frac{6G_\rho}{1+2F}
+\frac{6G_\rho +4\rho G_{\rho\rho}}{1+2F+4\rho F_\rho}\right.\nonumber\\
 & &\left.\qquad +8\rho G\left[\frac{G^2 F_\rho -(1+2F)(G+\rho 
G_\rho )G_\rho }{(1+\rho G^2)(1+2F+4\rho F_\rho )(1+2F)}\right]\right]
-2\rho G_\rho \,.
\eea
It is straightforward to check that these equations reproduce the 
perturbative running of the ordinary Yukawa and scalar quartic coupling in the 
absence of anomalous dimensions.
 
To proceed with the analysis, the WRGE must be supplemented with the form of 
the generalized potential at some initial scale $\Lambda = \Lambda (t=0)$. 
Choosing the $SU(2)_L \times U(1)$ invariant standard model form 
\be
U(\rho, Y; t=0) = \frac{1}{2} m^2(0) \rho + \frac{1}{4} \lambda(0) \rho^2 + 
\frac{1}{\sqrt{2}} g_t(0) Y \, ,
\ee
we then seek solutions for $F$ and $G$ which 
spontaneously break this symmetry, ie. $<\varphi_0> \not= 0$. This in turn 
restricts the $m^2(0), \lambda(0),g_t(0)$ parameter space.  
To implement the restriction, we define $m^2_{cr}$ as the maximum value of 
$m^2(0)$ for a given $\lambda(0), g_t(0)$ which results in a nontrivial zero 
of $F(\rho,t)$ as $t$ increases into the infrared. $m^2_{cr}$ is determined by 
evaluating $F(\rho,t)$ for $m^2(0)$ well into the broken phase and then 
increasing $m^2(0)$ until the zero of $F$ decreases as $t$ increases. The 
value of $m^2(0)$ which produces this transition so that the only zero of $F$ 
as $t\rightarrow\infty$ is at the origin defines $m^2_{cr}$. So doing, 
we establish 
the infrared Gaussian fixed point. From the solution, we also see that the 
induced irrelevant operators can give sizeable contributions for small $t$ 
values and play an important role in driving the theory toward this fixed 
point. Furthermore, the numerically generated solution to the WRGE is seen to 
smoothly join onto the 1-loop solution (which includes irrelevant operators) 
for $t$ values beyond some $t^*$.

To extract the Higgs scalar and top quark mass bounds, we choose a point in 
the allowed 
$\lambda(0), g_t(0), m^2(0) < m^2_{cr}$ parameter space corresponding to the 
spontaneous symmetry breaking solution and then numerically integrate the WRGE 
to 
obtain $F$ and $G$ for $0<t<t^*$. We next include the degrees of freedom with 
$\vert p \vert < e^{-t^*} \Lambda$ via the 1-loop perturbative solution which 
smoothly joins onto the numerically generated solution at $t^*$.  
As a consequence of the spontaneous symmetry breakdown, the model spectrum 
also contains three Nambu-Goldstone degrees of freedom which upon gauging the 
$SU(2) \times U(1)$ symmetry become, via the Higgs mechanism, the longitudnal 
components of the $W_{\pm}, Z$ vector bosons. Due to the presence of these 
massless 
modes, care must be exhibited in integrating all the way down into the 
infrared.  
Following the procedure used in the pure scalar model$^{[4]}$, we simply cut 
off the 
1-loop integrals in the infrared at a scale $p^2 = M_W^2$.  The computed 
masses turn out to be only weakly dependent on the choice of the infrared 
cutoff.  Summing the 
contributions produces the functions $F^{eff}(\rho)$ and $G^{eff}(\rho)$
from which the nontrivial scalar vacuum expectation value ($v$) and Higgs 
scalar and top quark masses ($M_H$ and $m_t$ respectively) are secured as
\bea
F^{eff} (\frac{v^2}{\Lambda^2})&=& 0\nonumber\\
\frac{M_H^2}{\Lambda^2} &=&  4 \frac{v^2}{\Lambda^2} F^{eff}_{\rho} 
(\frac{v^2}
{\Lambda^2})\nonumber\\
\frac{m_t}{\Lambda} &=& \left. G^{eff}(\frac{v^2}{\Lambda^2}) 
\frac{\partial Y}{\partial (\bar{t} t)}\right |_{\varphi_0 = 
\varphi_0^{\dagger} = v/(\sqrt{2}\Lambda)} \, .
\eea

\begin{center}
\let\picnaturalsize=N
\def\picsize{5.0in}
\def\picfilename{top6.ps}
\ifx\nopictures Y\else{\ifx\epsfloaded Y\else\input epsf \fi
\let\epsfloaded=Y
\centerline{\ifx\picnaturalsize N\epsfxsize \picsize\fi \epsfbox{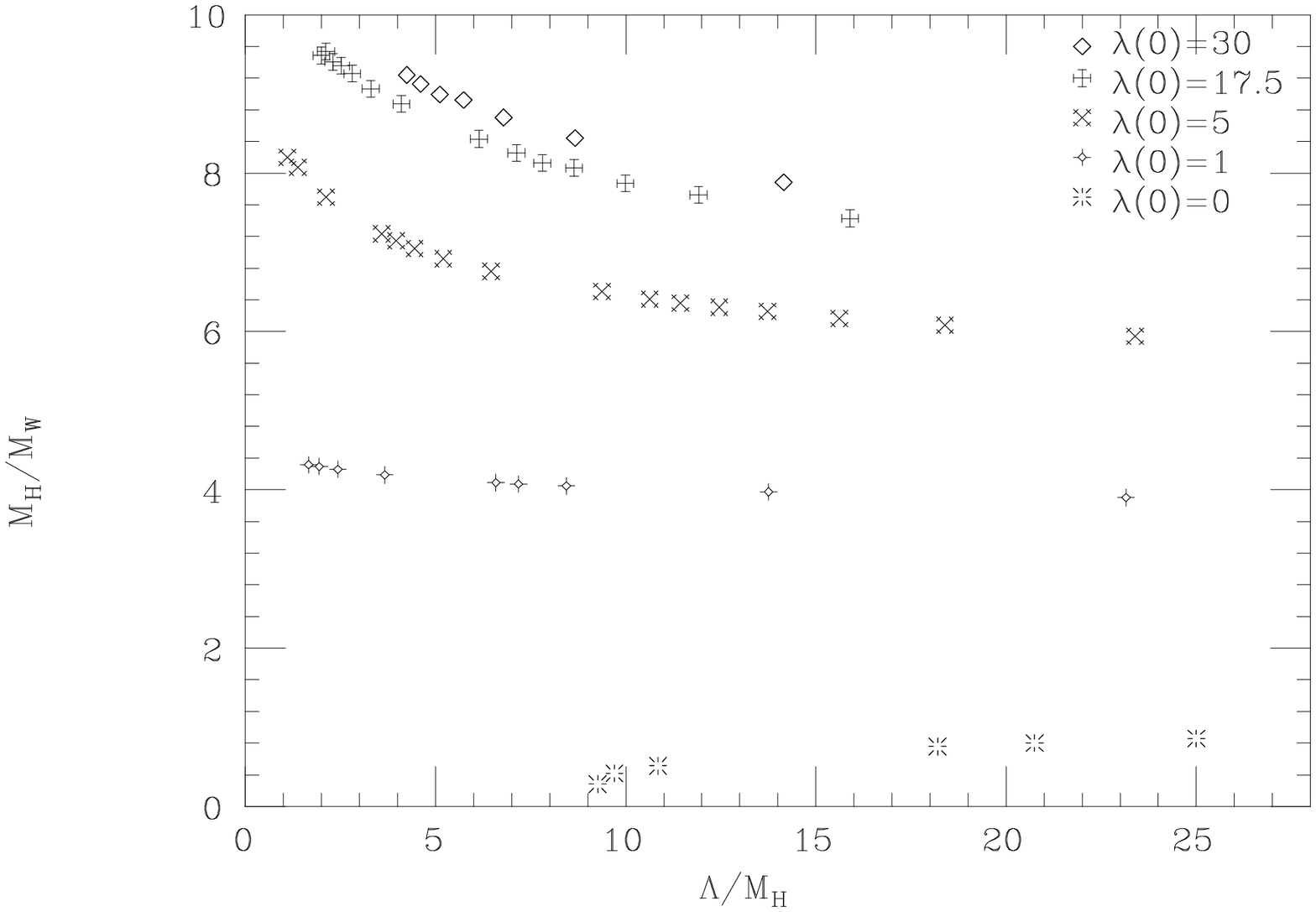}}}\fi
\medskip
{\sl Figure 1: $M_H/M_W$ as a function of $\Lambda/M_W$ for various $\lambda(0)$ 
values and $g_t(0)=0.6$.}
\end{center}

The value of $t^*$ is chosen such that the computed masses are stable to 
within an error the size of the $t$-grid spacing: $10^{-5}<\delta t< 10^{-4}$.  
Finally, the computation is 
repeated varying over the allowed parameter space. 

Using the above procedure, we extract the ratios $M_H/\Lambda, m_t/\Lambda$ and 
$v/\Lambda$ as the couplings $\lambda(0)$ and $g_t(0)$ are varied.  Taking 
various 
products of these ratios and using the relation between $v\simeq 246$ GeV 
and the $W_\pm$ mass, we plot in Figures 1 and 2 
the allowed $M_H/M_W$ values as a function of $\Lambda/M_H$ for a range of 
initial $\lambda(0)$ values and $g_t(0) = 0.6$ and $2.0$ respectively.  
The $g_t(0) = 0.6$ plot is virtually identical (to within the error bars) to 
the analogous $g_t(0) = 0$ plot$^{[4]}$.

\begin{center}
\let\picnaturalsize=N
\def\picsize{5.0in}
\def\picfilename{top2.ps}
\ifx\nopictures Y\else{\ifx\epsfloaded Y\else\input epsf \fi
\let\epsfloaded=Y
\centerline{\ifx\picnaturalsize N\epsfxsize \picsize\fi \epsfbox{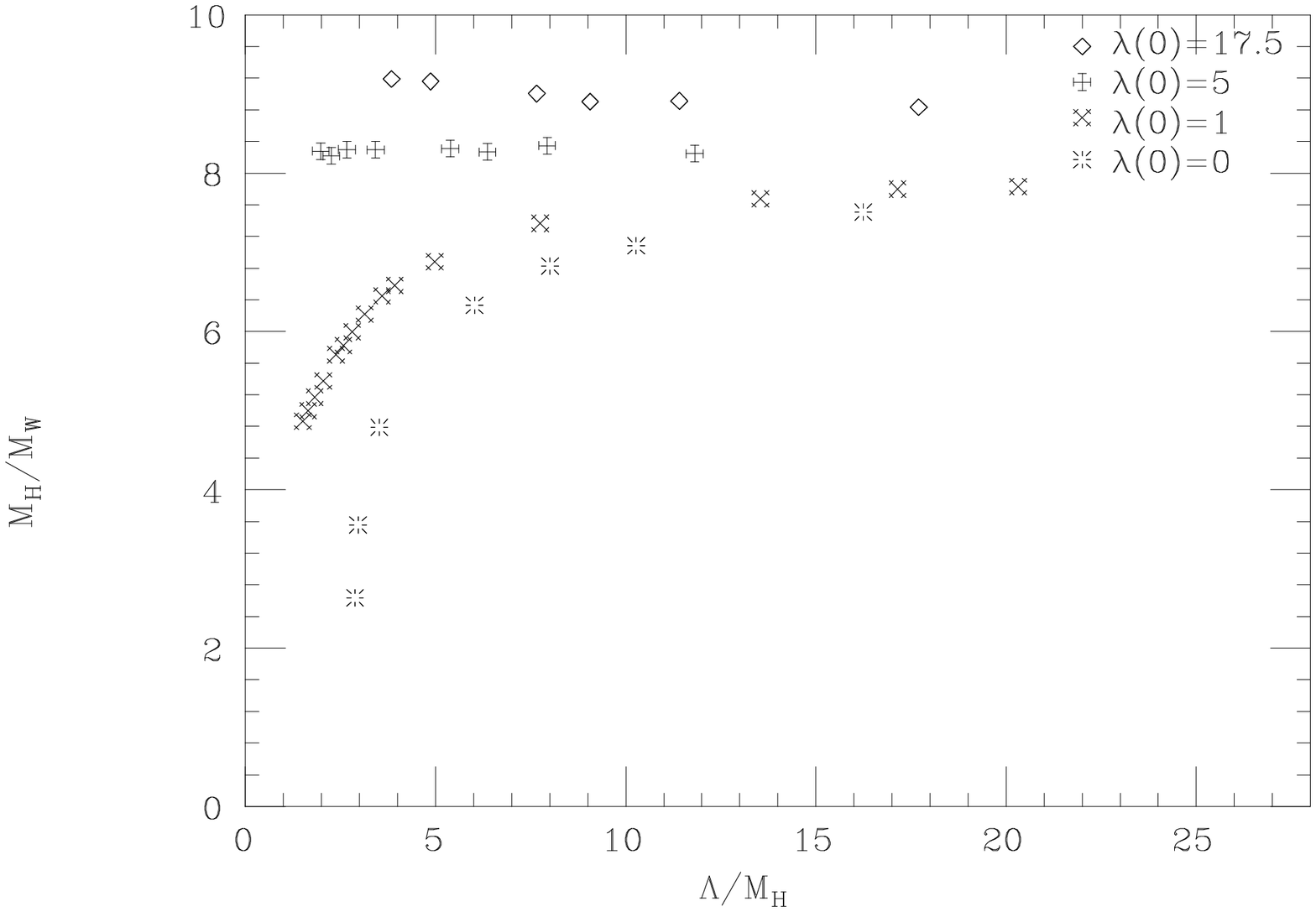}}}\fi
\medskip
{\sl Figure 2: $M_H/M_W$ as a function of $\Lambda/M_W$ for various $\lambda(0)$ 
values and $g_t(0)=2.0$.}
\end{center}

From the plotted data, we see in both cases that 
a dynamical envelope is formed as the initial $\lambda(0)$  
coupling is varied. That is, for fixed $\Lambda/M_W$, the different $M_H/M_W$ 
values appear to converge to an upper limit.  When this is combined with the 
physical requirement that no mass is allowed to become larger than 
the initial cutoff $\Lambda$ so that $\Lambda/M_H > 1$, we secure an upper 
bound on the allowed $M_H/M_W$ ratio of roughly $10$ which corresponds to 
the bound $M_H \leq 800$~GeV.  This is consistent with the results obtained 
using lattice simulations$^{[7]}$ as well as with the  
1/N$^{[8]}$ calculations.  Note that changing $g_t(0)$ from 0.6 to 2.0 produces 
but a small increase in the Higgs scalar mass upper bound.

We can also use the data to extract the allowed domain of scalar Higgs and top 
quark masses at a fixed cutoff to $M_W$ ratio. Figure 3 is just such
\begin{center}
\let\picnaturalsize=N
\def\picsize{5.0in}
\def\picfilename{toph.ps}
\ifx\nopictures Y\else{\ifx\epsfloaded Y\else\input epsf \fi
\let\epsfloaded=Y
\centerline{\ifx\picnaturalsize N\epsfxsize \picsize\fi \epsfbox{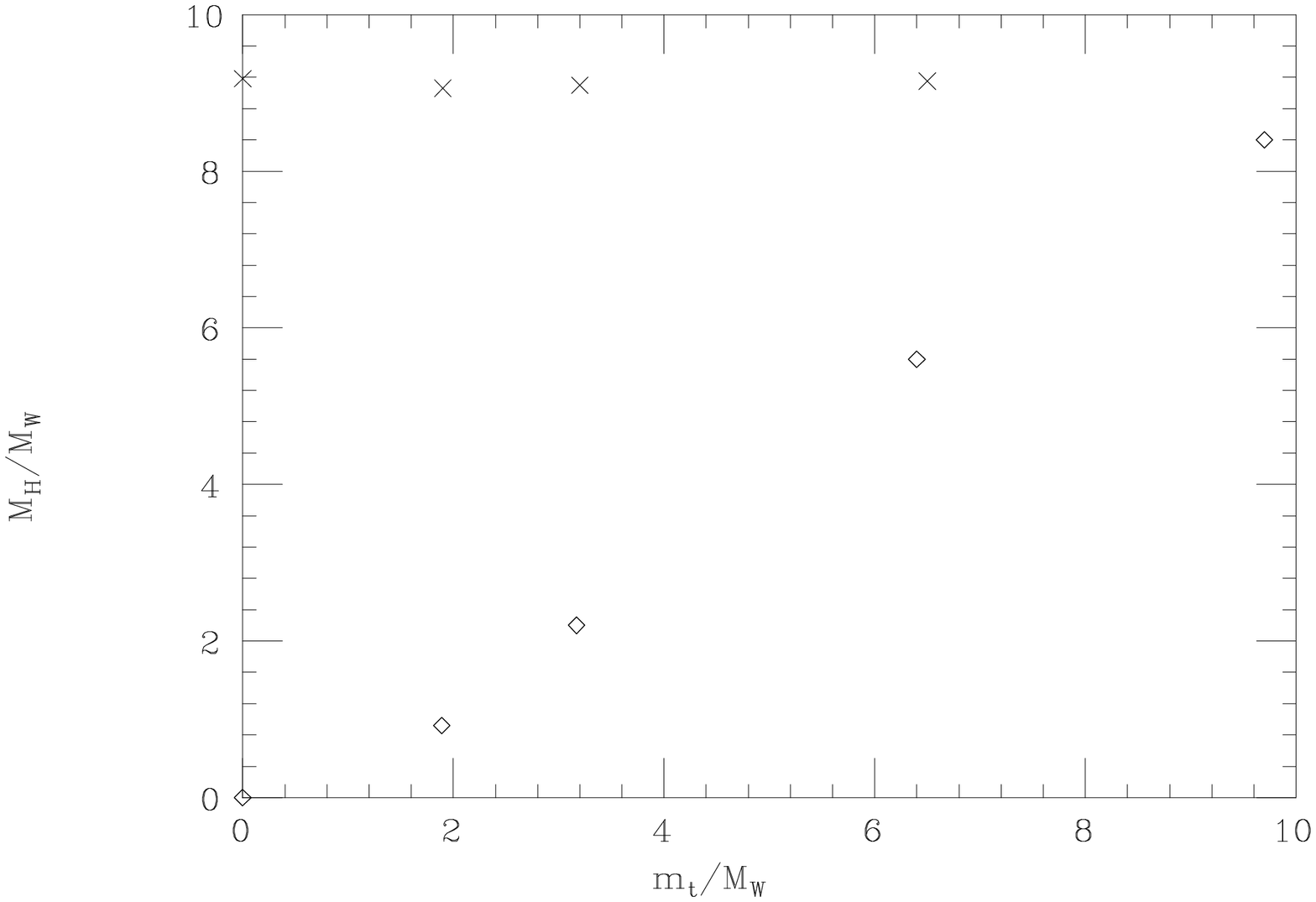}}}\fi
{\sl Figure 3: The allowed range of $M_H/M_W$ and $m_t/M_W$ values for 
$\Lambda/M_W=20$.}
\end{center}
a plot corresponding to $\Lambda/M_W = 20$.  The allowed values are those lying 
interior to the various boundaries displayed.  The upper portion of boundary, 
denoted by the cross ($\times$) marks is the scalar triviality bound discussed 
above.  
The flatness of this curve as a function of the top quark mass testifies to the 
insensitivity of the Higgs scalar mass upper bound to the top Yukawa coupling.  
The points on the figure represented by the diamond ($\diamond$) markings 
arise from the vacuum stability requirement which corresponds to the condition 
that $\lim_{|\rho|\rightarrow \infty} F(\rho, t) > 0$ for all $t$.  Note 
that for 
$m_t \simeq 200$ GeV, the lower bound on $M_H$ is around $60$ GeV which is 
roughly the current experimental lower bound.

We have used the Wilson renormalization group equation to construct 
nonperturbative mass bounds on the Higgs scalar and top quark degrees of 
freedom in the electroweak standard model.  We found a Higgs scalar mass 
absolute upper bound of approximately $800$~GeV which was quite insensitive to 
the presence of the top quark Yukawa coupling and thus basically identical to 
the WRGE calculation in the pure $O(4)$ scalar model$^{[4]}$ and a model with 
an additional scalar singlet$^{[9]}$.  Our results are also 
consistent with other mass bound estimates based on lattice 
simulations.  Of course, our study 
is subject to the various uncertainties which accompany the approximation 
schemes.  In particular, it is desirable to extend our analysis beyond the 
local approximation to include non-trivial anomalous dimensions and derivative 
interactions.  In addition, one should also include the effects of the gauge 
interactions and the light quark Yukawa couplings. Since these effects can 
be treated perturbatively$^{[10]}$ and are small we do not include them here.

\bigskip
 
\noindent
{\bf Acknowledgements}

The work of T.E.C., B.H. and S.T.L. was supported in part by the U.S. Department 
of Energy under grant DE-ACO2-76ER01428 (Task B).  The work of W.T.A.t.V. was 
supported by the U.S. Department of Energy under grant DE-FG05-8SER40226.

\newpage
\noindent
{\bf References}

\begin{enumerate}

\item
R. Dashen and H. Neuberger, Phys. Rev. Lett. {\bf 50} (1983) 1897; M.A.B. Beg 
et al, Phys. Rev. Lett. {\bf 52} (1984) 833; D.J.E. Callaway, Nucl. Phys. 
{\bf B223} (1984) 189; M. Lindner, Z. Phys. {\bf C31} (1986) 295. 

\item
M. L\"uscher and P. Weisz, Phys. Lett. {\bf 212B} (1988) 472; Nucl. Phys. 
{\bf B318} (1989) 705; J. Kuti, L. Lin and Y. Shen, Phys. Rev. Lett. {\bf 61} 
(1988) 678; G. Bhanot, K. Bitar, U. Heller and H. Neuberger, Nucl. Phys. 
{\bf B353} (1991) 551.

\item
M. Einhorn, Nucl. Phys. {\bf B246} (1984) 75.

\item
P. Hasenfratz and J. Nager, Z. Phys.{\bf C37} (1988) 477.

\item
K.G. Wilson and J.B. Kogut, Phys. Rept. {\bf 12C} (1974) 75; 
K.G. Wilson, Rev. Mod Phys. {\bf 55} (1983) 583; 
F.J. Wegner and A. Houghton, Phys. Rev. {\bf A8} (1973) 401;
S. Weinberg, in {\bf Proceedings of the 1976 International School of 
Subnuclear Physics}, Erice, ed. by A. Zichichi (Plenum Press, 1978); 
see also J. Polchinski, Nucl. Phys. {\bf B231} (1984) 269; C. Wetterich, Nucl.
Phys. {\bf B352} (1991) 529; U. Ellwanger and L. 
Vergara, Nucl. Phys. {\bf B398} (1993) 52; T. Morris, preprint 
CERN-TH.6977/93.

\item
T.E. Clark, B. Haeri and S.T. Love, Nucl. Phys {\bf B402} (1993) 628.

\item
Y. Shen, J. Kuti, L. Lin and P. Rossi, Nucl. Phys. (Proc. Suppl.) {\bf 9} 
(1989) 99.

\item
M. Einhorn and G. Goldberg, Phys. Rev. Lett. {\bf 57} (1986) 2115; G. Bathas 
and H. Neuberger, Nucl. Phys. (Proc. Suppl.) {\bf 30} (1993) 635.

\item
R. Akhoury and B. Haeri, Phys. Rev. {\bf D48} (1993) 1252.

\item
L. Maiani, G. Parisi and R. Petronzio, Nucl. Phys. {\bf B136} (1978) 115; 
N. Cabibbo et al, Nucl. Phys. {\bf B158} (1979) 295.
\end{enumerate}
\end{document}